# Coulomb blockade based field-effect transistors exploiting stripe-shaped channel geometries of self-assembled metal nanoparticles

*Hauke Lehmann\*, Svenja Willing, Sandra Möller, Mirjam Volkmann, and Christian Klinke\**

University of Hamburg, Institute of Physical Chemistry,
Grindelallee 117, 20146 Hamburg, Germany

E-mail: lehmann@chemie.uni-hamburg.de, klinke@chemie.uni-hamburg.de

**Abstract**

Metallic nanoparticles offer possibilities to build basic electric devices with new functionality and improved performance. Due to the small volume and the resulting low self-capacitance, each single nanoparticle exhibits a high charging energy. Thus, a Coulomb-energy gap emerges during transport experiments that can be shifted by electric fields, allowing for charge transport whenever energy levels of neighboring particles match. Hence, the state of the device changes sequentially between conducting and non-conducting instead of just one transition from conducting to pinch-off as in semiconductors. To exploit this behavior for field-effect transistors, it is necessary to use uniform nanoparticles in ordered arrays separated by well-defined tunnel barriers. In this work, CoPt nanoparticles with a narrow size distribution are synthesized by colloidal chemistry. These particles are deposited via the scalable Langmuir-Blodgett technique as ordered, homogeneous monolayers onto Si/SiO$_2$ substrates with pre-patterned gold electrodes. The resulting nanoparticle arrays are limited to stripes of adjustable lengths and widths. In such a defined channel with a limited number of conduction paths the current can be controlled precisely by a gate voltage. Clearly pronounced Coulomb oscillations are observed up to temperatures of 150 K. Using such systems as field-effect transistors yields unprecedented oscillating current modulations with on/off-ratios of around 70%.

**Introduction**

Nowadays, integrated circuits contain over a billion transistors based on standard silicon technology.[1] However, the conventional down-scaling is about to reach its physical limits.[2] Short-channel effects, that is basically the reduction of the gate influence on the channels that decrease in size, significant leakage currents, and a reduction of the threshold voltage become major problems. To continue the miniaturization without loss of performance and to obey Moore's Law new designs, concepts, or materials for transistors have to be explored to allow for smaller, faster, cheaper, and at the same time reliable and more energy efficient devices.[3] Transistors based on the Coulomb blockade represent such a possible future device concept. Although they might not be able to compete with standard silicon technology at the moment, they offer three main advantages: (1) they can be down-scaled to almost arbitrary size (ultimately down to one individual nanoparticle), (2) transistors based on small particle arrays can be fabricated easily and inexpensively as demonstrated by the approach in this paper and most of all, (3) they exhibit a novel oscillating transfer characteristic that might be useful far beyond the mere transistor application. In general, the Coulomb blockade regime can be overcome by applying a sufficient bias voltage to induce tunneling through the barriers or by thermally activated electron hopping. For a lithographically defined tunnel-junction this has already been demonstrated several years ago.[4-6] In single-electron transistors the application of a gate voltage manipulates the energy levels of an island and adjusts its number of electrons. This was also demonstrated for one or a few individual nanoparticles.[7-11] While general electrical transport properties of granular nano-materials have been studied extensively,[12-18] there are not many reports on two-dimensional metal particle systems, especially including the influence of an external electric field. It has recently been shown that extended metal-nanoparticle films can be used as Coulomb blockade based field-effect transistors.[19] However, in order for them to present a real alternative for industrial purposes, their performance and the working temperature range has to be improved. Therefore, a deeper understanding of the predominant charge-transfer mechanisms as well as the influencing parameters in each device is mandatory. This knowledge can in turn be used to optimize the current flow and the field effect and thereby to increase the efficiency of the improved devices. Investigating the basic physics in such a device as well as analyzing the influences of different parameters and their direct or indirect effect on the performance is on the one hand interesting for the application itself. However, on the other hand, the metal particle array is of general scientific interest due to the fact that metals are usually not susceptible to a gate voltage. Thus, dependencies of the current-voltage characteristics on different system parameters should be studied.

One of the advantages of nanoparticles is that they are tunable in size, which allows for precise adjustment of their electrical and optical properties.[20,21] An individual conductive spherical particle with dimensions in the nanometer range possesses a small self-capacitance of

$$C = 4 \cdot \pi \cdot \varepsilon_0 \cdot \varepsilon_r \cdot r + C_{surrounding} \quad (1)$$

that is proportional to the particle's diameter $r$ and depends on the relative permittivity $\varepsilon_r$ of the dielectric medium. When including such a particle into a current path, also the capacitance of the surrounding geometry has to be added as $C_{surrounding}$. However, this is negligible compared to the self-capacitance of the particle.[22] Charging the particle with an additional electron requires the energy

$$E_C = \frac{e^2}{2 \cdot C} \quad (2)$$

due to Coulomb repulsion.[23,24] If the thermal energy of the electron is not sufficient to exceed the charging energy $E_C$, this leads to a virtually insulating character of the structures known as Coulomb blockade. The blockade can be overcome either by increasing the temperature to generate electron hopping or by applying a bias voltage between the source and drain electrodes, which is large enough that electrons start to tunnel through the barrier resulting in a net current.[4-6] In addition to the energy of the electrons provided by temperature or a lateral electric field, the transport mechanism and the current through the system depend on various other parameters. The charging energy is influenced by the particle size and the dielectric constant of the surrounding medium (compare Equation 1 and 2). While the spacing of the resulting discrete energy levels plays an important role in the tunneling rate, the energy-level distribution, the order of the system, and the inter-particle distance determine whether it is more favorable to tunnel to the adjacent particle according to nearest-neighbor hopping,[24,25] or over varying distances as described by variable-range hopping.[26-28]

To examine the transfer characteristics in a three-terminal device, a gate electrode is added. An electric field, created by the application of a gate voltage, presents an easy accessible possibility to externally influence the system. Tuning the gate voltage shifts the energy levels of the island and thereby adjusts the number of electrons on this island.[7-9] The position of the first unoccupied level in an island with respect to the Fermi-energy of neighboring particles or the electrodes strongly influences the probability of tunneling. Thus, a continuous change in gate voltage will lead to periodic peaks of the current flow whenever the energy levels match. Each current maximum corresponds to a charge difference of one electron on each particle at the average. This characteristic behavior makes it possible to use such a system as a transistor device with successive on- and off-states. The Coulomb-energy gap then adopts the role of a semiconductor bandgap. However, unlike in a conventional field-effect transistor with just one transition from conducting to pinch-off, the unique transfer characteristics with periodic on- and off-states might allow for new concepts and possible applications.

We synthesize CoPt-nanoparticles with a diameter of $(4.0 \pm 0.3)$ nm by colloidal chemistry. These particles are stable against oxidation and electromigration. In addition, they are tunable in size, shape, and ligand length. Their solution-processability makes them easy to handle and their ambition to self-assemble allows for the scalable deposition of ordered, homogeneous monolayers of nanoparticles with narrow size distribution onto various substrates via the Langmuir-Blodgett technique.[17,19,22,29-33] The organic ligands precisely define the tunnel-barrier widths and keep the nanoparticles separated, as a merging would render the system completely metallic. Since the particles possess a narrow size distribution ($< 8\%$), all of them are simultaneously in the state of Coulomb blockade (a higher degree of polydispersity of the nanoparticles would smear out the energy levels and deteriorate the function as transistor). Due to the free out-of-plane direction the monolayer is well accessible for the gate electrode. The energy levels can be influenced efficiently and simultaneously by the electric field and screening is hindered.

Pre-defined gold electrodes, a resist mask with stripes of defined lengths and widths, and the subsequent preparation of a top-gate electrode enable the fabrication of a precisely controllable device that provides transistor functionality. The feasibility of metal nanoparticle based field-effect transistors has previously been proven with larger CoPt nanoparticles of $(7.6 \pm 0.5)$ nm in diameter using vast films.[19] The performance regarding on/off-ratio and maximum working temperature is now greatly improved through the implementation of the channel as a two-dimensional nanoparticle array/stripe of limited size. Restricting the film to such stripes allows for two main advantages: On the one hand, the number of possible conduction paths and thereby the current through the system can be tuned to fulfill requirements of a possible later application. On the other hand, the top-gate electrode is now able to completely cover and thereby efficiently influence all conduction paths. As a consequence of the improved device geometry resulting in a smaller area of the nanoparticle film and improved electrostatic control, also smaller leakage currents are ensured while each device is single-addressable. Furthermore, our improvements include the use of smaller particles resulting in an increase of the charging energy. Thus, the maximum working temperature, at which Coulomb oscillations can still be observed, is increased.

Output and transfer characteristics are investigated on various sample geometries and at different measurement temperatures to examine Coulomb blockade and Coulomb oscillations. We observe clearly pronounced Coulomb oscillations above liquid-nitrogen temperature and even up to 150 K and demonstrate that on/off-ratios in the range of 70% are possible in our devices. This success is attributed to the high quality of the small and well-organized CoPt metal nanoparticles in the respective stripes.

**Experimental**

Spherical CoPt nanoparticles with a diameter of (4.0 ± 0.3) nm are synthesized by colloidal chemistry according to an established approach.[34] The platinum source, platinum(II) acetylacetonate ($Pt(acac)_2$, 0.08 mmol), is dissolved in hexadecane and will be reduced to metallic platinum during the synthesis. The ligands, decylamine (12 mmol) and decanoic acid (1.4 mmol), that control shape and size during particle nucleation and growth and then act as stabilizing agents to keep the nanoparticles soluble, are also included into the mixture. The solution is heated to 80 °C and stirred at this temperature for at least an hour under vacuum conditions at a pressure of $5.3 \cdot 10^{-2}$ mbar. The cobalt source, dicobalt octocarbonyl ($Co_2(CO)_8$, 0.12 mmol), dissolved in 1,2-dichlorobenzene in an ultrasound bath, is injected rapidly into the solution that has been heated to 160 °C under nitrogen atmosphere. This so-called hot-injection [35-37] is followed by thermal decomposition of the cobalt precursor. The mixture immediately turns black and is continuously stirred at the injection temperature for two hours. The obtained particles are precipitated with methanol and 2-propanol, centrifuged, and subsequently re-suspended in toluene to keep them solution-processable over long periods of time. Applying Vegard's law [38] to X-ray diffraction measurements reveals a nanoparticle composition of 88% Pt and 12% Co in the final product.

Highly doped silicon wafers with a 300 nm thick capping layer of thermal oxide are used as substrates. Standard electron-beam lithography with positive resist, PMMA (Poly(methyl-methacrylate)), is employed to fabricate the mask for the source- and drain-electrode geometry. Thermal evaporation of 2 nm titanium as adhesion layer and 23 nm gold for the leads and contact pads and a subsequent lift-off process in acetone transfers the geometry onto the chip. Stripes of different lengths and widths between source and drain are prepared by another electron-beam lithography step using the same resist. For precise adjustment of the stripes, a marker-alignment procedure is used. The markers have been prepared together with the gold electrodes.

The Langmuir-Blodgett method [29,30] is employed to fabricate ordered, homogeneous, and scalable monolayers of CoPt nanoparticles.[17] Due to the resist mask the particles only reach the surface of the sample in the resist-free stripe regions. Prior to film preparation the particles are precipitated with methanol and centrifuged ("washed") two more times, re-suspended in toluene, and spread onto a subphase of diethylene glycol. After the evaporation of the solvent, the nanoparticle film is slowly compressed by two barriers with a speed of 1 mm/min to a target pressure that is approximately 7 mN/m higher than the initial surface pressure.[39] The respective target pressure is held for 2 h, allowing for the film to relax and for the particles to rearrange. A transmission-electron microscope (TEM) image of the resulting homogeneous monolayer on a carbon-film coated copper grid can be seen in **Fig. 1(a)**. Lift-off in acetone removes the resist together with the nanoparticles on top of the mask leaving only the previously uncovered stripes between source and drain now loaded with nanoparticles. These nanoparticles stick to the substrate due to the relatively strong van-der-Waals forces at the nanoscale. The resulting stripes are then protected by sputter deposition of 150 nm silicon-dioxide ($SiO_2$). This insulating layer serves as dielectric between channel and top-gate electrode, which is carefully centered in the middle of the channel. It is prepared by another electron-beam lithography step, thermal evaporation of 2 nm titanium as adhesion layer and 38 nm gold, and a subsequent lift-off process. The use of local top-gate electrodes instead of applying a global back-gate voltage to the whole sample makes it possible to address single devices individually. An optical micrograph of a device is included in **Fig. 1(b)**, while a cross-sectional scheme is given in **Fig. 1(c)**. The devices are electrically characterized by DC-measurements in a high vacuum of around $10^{-6}$ mbar at temperatures between 8 K and room temperature (RT) in a closed-cycle probe station.

**Results and discussion**

Different geometries of metal-nanoparticle stripes with lengths between 1 µm and 10 µm and widths between 0.25 µm and 5 µm have been fabricated and electrically characterized. Following bulk resistor expectations, the nanoparticle stripes that are shortest and widest result in the highest currents due to a limited number of tunnel barriers to cross and various possible conduction paths in parallel. Tuning the geometry of the stripe allows for current adjustment over some orders of magnitude. However, it has to be kept in mind that the gate influence decreases if charge carriers can travel along numerous paths. Thus, a compromise between signal-to-noise ratio and controllability has to be made. The electronic noise is due to statistical and thermal variations in the tunneling process of the electrons and is obviously most pronounced in low current measurements.

Output characteristics of all investigated devices show the transition from a nearly Ohmic behavior (as expected for a metallic system) at room temperature to a system governed by Coulomb blockade at temperatures below a certain critical temperature. The absolute current can be adjusted by the stripe geometry and the source-drain voltage. The threshold voltage, that defines the onset of electrical transport in terms of measurable current flow, as well as the critical temperature and in general the quality of the signal with respect to electronic noise varies from device to device depending on the respective geometry. Stripes with short channel lengths and wide channel widths yield high absolute currents, a small threshold voltage, and a good signal-to-noise ratio even at low temperatures. In contrast, stripes with long channel lengths and narrow channel widths exhibit high resistances resulting in small currents through the device and a wide Coulomb-blockade regime with a high threshold voltage. In the following, results from representative devices are shown exemplarily and general trends are discussed. The threshold voltage and critical temperature define the point at which the Coulomb blockade can be overcome by a sufficiently high bias voltage or thermal activation respectively. However, it has to be kept in mind, that both quantities are determined by analyzing the onset of electrical current flow or appearance of a finite conductance which completely depends on the sensitivity limit of the measurement setup. Nevertheless, it is common to define these limits in order to describe the data.

**Fig. 2(a, c)** shows current-voltage curves obtained from two stripes of 2 µm length and 0.5 µm or 2 µm width respectively at different temperatures. At low temperatures the current is small and the current-voltage curve is clearly S-shaped. However, it is worth mentioning

that also at room temperature the current-voltage curve is not a linear Ohmic slope like one would expect for a bulk metal. Instead, the tunnel barriers between the nanoparticles still partially sustain the non-Ohmic character and render the curve slightly S-shaped. To determine the temperature-dependent threshold voltages and for better visibility of the Coulomb-blockade regime, the obtained output characteristics of the two devices are displayed in **Fig. 2(b, d)** as differential conductances on the logarithmic scale. With increasing temperature first the width of the Coulomb-blockade regime decreases until a finite zero-bias conductance appears. The critical temperature for this transition from vanishing to finite zero-bias conductance strongly depends on the number and length of conduction paths and, thus, on the overall resistance of the respective device. For the investigated devices it ranges between 28 K and 78 K, being highest for long and/or narrow stripes where the conductance is low at any temperature.

Predominant transport mechanisms are identified by fitting the data at lowest and at room temperature according to different models. **Fig. 3(a)** indicates that in the temperature regime of 8 K tunnel processes dominate the transport as described with the Nordheim-Fowler relation.[40] The current

$$I \propto V^2 \cdot \exp\left(\frac{-B}{V_{sd}}\right) \qquad (3)$$

with the fitting parameter $B$ strongly depends on the source-drain voltage $V_{sd}$, is basically temperature independent, and can result from direct or field-induced tunneling.[41,42]

By using the temperature dependency instead of the current-voltage curves it is possible to analyze the transport behavior further. The differential conductance can in general be described by

$$G \propto \exp\left(\left(\frac{T_0}{T}\right)^\nu\right) \qquad (4)$$

with a constant $T_0$ and the critical exponent $\nu$ ranging between zero and unity. Typical processes for our system include tunneling between neighboring sites with its Arrhenius behavior of $v = 1$.[24,25] Another possibility is variable-range hopping (VRH),[26,27] which depends on the system's dimension $D$ and yields $v = (D+1)^{-1}$, such that $v = 1/3$ in two-dimensional films. VRH occurs, when it is most favorable for electrons to tunnel elastically to a more distant nanoparticle which has an energy level very similar to the initial state. If Coulomb interactions play an important role, the Efros-Shklovskii extension to VRH gives a dimension independent exponent of $v = 1/2$.[24,28] The same temperature behavior was suggested for co-tunneling, a transport mechanisms with two simultaneous tunnel processes.[16] Fitting the zero-bias conductance at temperatures above the Coulomb blockade with the three exponents reveals that the data can best be described with $v = 1$, suggesting nearest-neighbor hopping which can be attributed to the narrow size distribution of the particles.[24,25] In this case, $T_0$ is given by the activation energy $E_A$ divided by the Bolzmann constant $k_B$. From the slope of the Arrhenius fits, a plot can be found in Ref. [22], an activation energy of $E_A = (26.8 \pm 0.4)$ meV has been determined, which is in the same order of magnitude as for comparable particles investigated previously.[22]

With increasing temperature thermal contributions influence the transport and increase the charge-carrier concentration. The current flow at room temperature can be attributed to different thermally induced processes like thermally activated tunneling, or thermionic emission.[41,42] A comparison of the fits for tunneling and thermal contributions at room temperature is given in **Fig. 3(b)**. With increasing temperature the pure tunneling is superimposed by increasing thermal contributions.

Local top-gate electrodes, which are carefully centered above the nanoparticle stripes and separated from the nanoparticle channel by a layer of silicon-dioxide (SiO$_2$), allow to investigate the transfer characteristics of our metal nanoparticle based field-effect transistors. The SiO$_2$ provides not only electrical insulation but, as a dielectric, it also increases the particles' self-capacitance leading to an even larger Coulomb-energy gap according to Equation 1 and 2. The layer thickness has to be carefully chosen to maximize the gate's field effect without risking leakage currents through the dielectric. The efficiency of the device, also known as relative on/off-ratio, is given by

$$R_{\%} = \left| \frac{I_{min} - I_{max}}{I_{max}} \right| \quad (5)$$

where $I_{min}$ and $I_{max}$ are the local extrema of the current in the analyzed gate-voltage regime.

Similar to previous experiments on vast films,[19] we observe Coulomb oscillations in the transfer characteristics of all analyzed devices. While this observation is valid for every investigated device, it varies in quality. Since only a fraction of the total current through the system can be influenced efficiently by a gate voltage, the oscillations on this tunnel current are superimposed by a constant current signal resulting from thermal contributions, which increases with increasing temperature. Hence, in devices with highly conducting short and wide stripes, the relative oscillation amplitude is small compared to the one measured in devices that conduct only near the noise level. Thus, the field effect is most visible if only a few electrons (resulting in a small current) flow preferably along just one conduction path without any parallel paths. In addition to the stripe geometry, also the applied source-drain voltage as well as the temperature determine the current through the system and, thus, the relative on/off-ratio.

Main features of the oscillations, like the position of the extrema, are expected to reproduce in different gate-voltage sweeps as well as for different bias voltages and temperatures. This is used to validate the occurrence of oscillations against interferences or noise. First, the temperature dependency of the on/off-ratio is examined. Maximum on/off-ratios of around 44% are obtained at a temperature of 8 K and with a source-drain voltage of -3 V, compare **Fig. 4**. These parameters ensure the best possible controllability since the thermally activated current is reduced to a minimum and the chosen source-drain voltage is just above the respective threshold, meaning there are only a small number of possible conduction paths (in the ideal case just one) to be influenced. With increasing temperature the threshold voltage decreases and the absolute current rises. At temperatures above the critical temperature for Coulomb blockade, no threshold voltage can be defined in the output characteristics and a small but non-vanishing constant bias voltage of -1 V is applied during the gate-voltage sweep. Also without a pronounced Coulomb-blockade regime in the output characteristics we observe Coulomb oscillations in the transfer characteristics, which is according to theory.[4x2] However, due to previous resolution limits of the setup, this has not yet been observed. While the absolute oscillation amplitude increases with temperature, the shrinking relative oscillation amplitude becomes harder to detect due to the growing current signal. Nevertheless, oscillations in this specific device under investigation are clearly observed up to temperatures of 150 K, but there the on/off-ratio ranges only in the region of some few percent. Above this temperature the total current through this nanoparticle stripe is too high to reliably identify the superimposed Coulomb oscillations.

At the lowest investigated temperature of 8 K, also the influence of the applied source-drain voltage was examined, as shown in **Fig. 5**. Coulomb oscillations have clearly been detected for all chosen source-drain voltages and show a good reproducibility of the key features. Even if the bias voltage is below the threshold voltage and the electrons do not have sufficient energy to overcome the barriers, a tuning of the energy levels inside the particles through an applied gate voltage may lead to electrical transport and also here Coulomb oscillations can be recorded by sweeping the gate voltage. Thus, the system can be periodically driven from a basically insulating state inside the Coulomb-blockade regime to a conducting state via the field-effect provided by the gate voltage. The resulting relative on/off-ratio is large due to the virtually vanishing current at zero gate voltage. Hence, the detected signal is more or less completely coming from the oscillatory behavior. This allows for an on/off-ratio of up to 70%, which is a much higher value than achieved before.[19] Although under these conditions the oscillations are quite noisy and the total current through the stripe is rather small, this proves that we succeeded to overcome the Coulomb blockade by means of a sufficient gate voltage. With further increasing source-drain voltage again the absolute oscillation amplitude grows. However, the relative oscillation amplitude superimposed to the increasing current signal is decreasing which makes the Coulomb oscillations more and more difficult to detect outside the Coulomb-blockade regime. Nevertheless, even at a bias voltage of -15 V, which is far outside of the blockade regime, the oscillations are still clearly visible and amount to a relative on/off-ratio of a few percent.

This behavior with increasing bias voltage is in good agreement with the temperature dependent results reported above. Both, rising temperature and growing applied bias voltage can lift the Coulomb blockade, leading to higher currents and a higher absolute oscillation amplitude, but also to lower relative on/off-ratios. However, in both cases we are still able to observe the superimposed Coulomb oscillations and to identify them as such by the reproducing features.

**Conclusions**

In this work, we demonstrate that well-defined stripes made of an ordered monolayer of CoPt nanoparticles with a diameter of $(4.0 \pm 0.3)$ nm are successfully fabricated by Langmuir-Blodgett deposition via resist masks on predefined gold electrodes. The ligand shells serve as tunnel barriers and suppress the metallic behavior. This is an easy and inexpensive preparation method to provide transistor functionality.

Improvements compared to the earlier experiments are due to two facts. In the improved transistors, the current paths are restricted to stripes. Hence, the top-gate electrode covers the whole width of the channel and all electrons in their current paths are influenced simultaneously by the gate voltage. Also, we successfully employed the Langmuir-Blodgett method for CoPt nanoparticles of only 4 nm in diameter. Thus, the Coulomb-energy gap increases and the occupation or depletion of individual energy levels is more reliably controlled by the gate voltage. Furthermore, the effects are better detectable and persistent up to higher temperatures overall resulting in larger on/off-ratios.

At room temperature, thermal contributions govern the transport while low temperature measurements are dominated by Coulomb blockade and give evidence for tunneling to be the predominant transport mechanism.

Tuning the energy levels of the nanoparticles with an electric field leads to the possibility of a current flow even if the energy of the electrons is insufficient to overcome the Coulomb-energy gap. At 8 K we observe transport in the blockade regime that leads to a high on/off-ratio of up to 70%. Transfer characteristics have been analyzed with respect to the applied bias voltage as well as to the measurement temperature. An increase in each of the two parameters leads to higher absolute oscillation amplitudes due to more possible conduction paths, additional charge carriers, or thermally activated transport. However, also the overall current through the system is raised and can only partially be influenced by the gate voltage. Thus, the relative oscillation amplitudes as well as the relative on/off-ratios decrease with higher temperatures or larger source-drain voltages. By choosing appropriate stripe geometries and source-drain voltages, the general conductance of the channel as well as the strength of the Coulomb oscillations can be adjusted. Whereas stripes with short channel lengths and wide channel widths yield high absolute currents followed by smooth periodic oscillations, a small threshold voltage, and a good signal-to-noise ratio even at low temperatures, the relative on/off-ratios of such devices are rather small. In contrast, stripes with long channel lengths and narrow channel widths exhibit high resistances resulting in small currents through the device, a wide Coulomb-blockade regime with a high threshold voltage, and high relative on/off-ratios due to the small superimposed current.

All in all, we have detected reliable and reproducible Coulomb oscillations up to temperatures of about 150 K which is well above the liquid-nitrogen temperature. The possibility to tune the system to match the requirements of a given application as well as the unique oscillating transfer characteristics makes the devices suitable to be used as field-effect transistor.

Compared to a conventional semiconductor field-effect transistor with just one transition from conducting to pinch-off, the most intriguing new accomplishment is the characteristic oscillating behavior. Such a device could be used for current stabilization if operated at a gate voltage that corresponds to one of the oscillation extrema. The possibility to drive the system through a sequence of on- and off-states by tuning the gate voltage might even be a unique and interesting feature for new device concepts that have to be developed. The fabrication of the proposed device can be done by applying an inexpensive, fast, and scalable preparation method using colloidal metal nanoparticles.

**Acknowledgements**

The authors would like to thank Sascha Kull for synthesizing the CoPt nanoparticles. Financial support of the European Research Council via the ERC Starting Grant "2D-SYNETRA" (Seventh Framework Program FP7, Project: 304980) as well as via the Heisenberg scholarship KL 1453/9-2 of the Deutsche Forschungsgemeinschaft (DFG) is gratefully acknowledged.

**Figures**

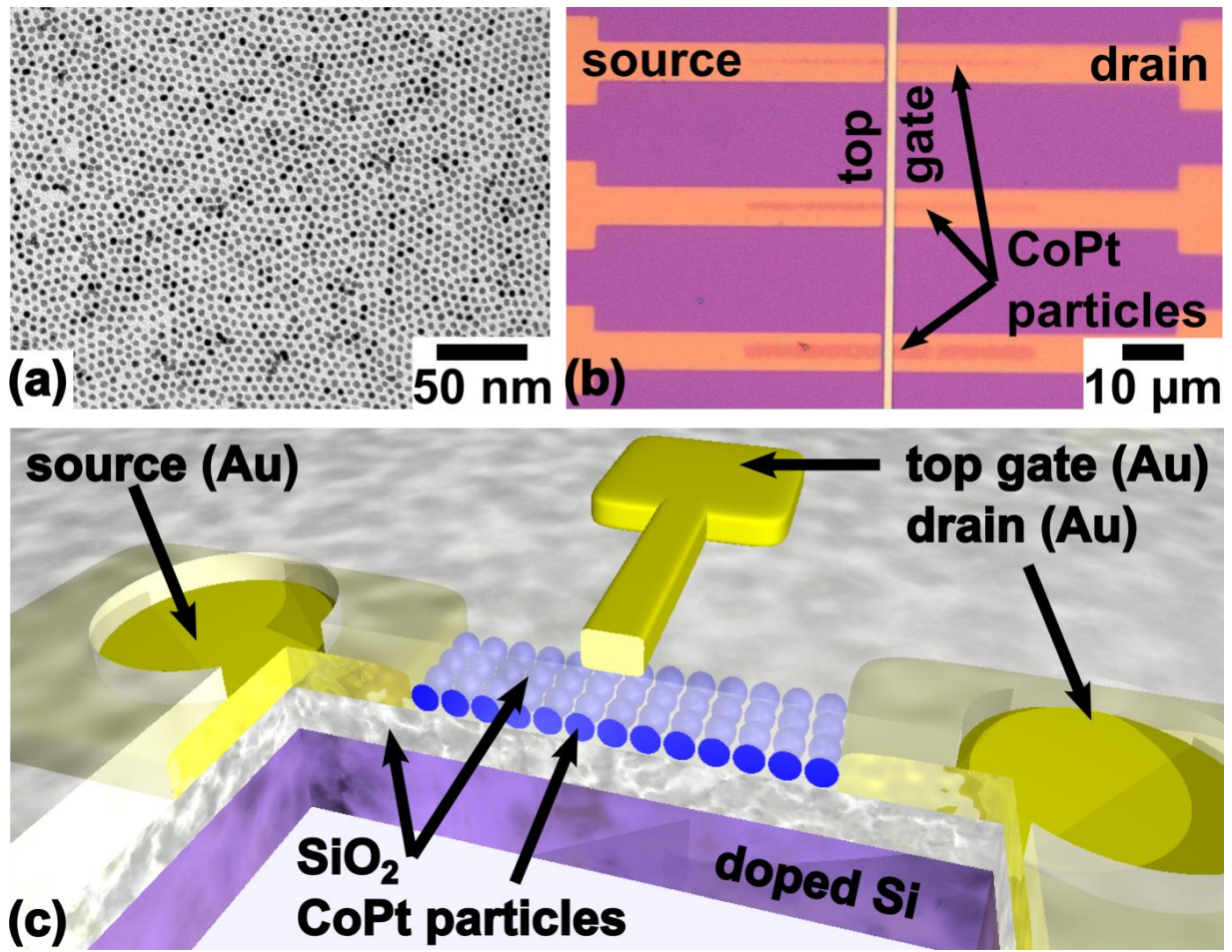

**Fig. 1** (a) Transmission-electron micrograph of a monolayer of spherical CoPt nanoparticles with a diameter of $(4.0 \pm 0.3)$ nm. The TEM grid has been loaded with nanoparticles by applying the Langmuir-Blodgett technique. (b) Optical micrograph of nanoparticle loaded stripes between gold electrodes. This sample has been fabricated in the same Langmuir-Blodgett process as the TEM grid in (a). The top-gate electrode is fabricated on a $SiO_2$ dielectric layer. (c) Cross-sectional scheme of the final device.

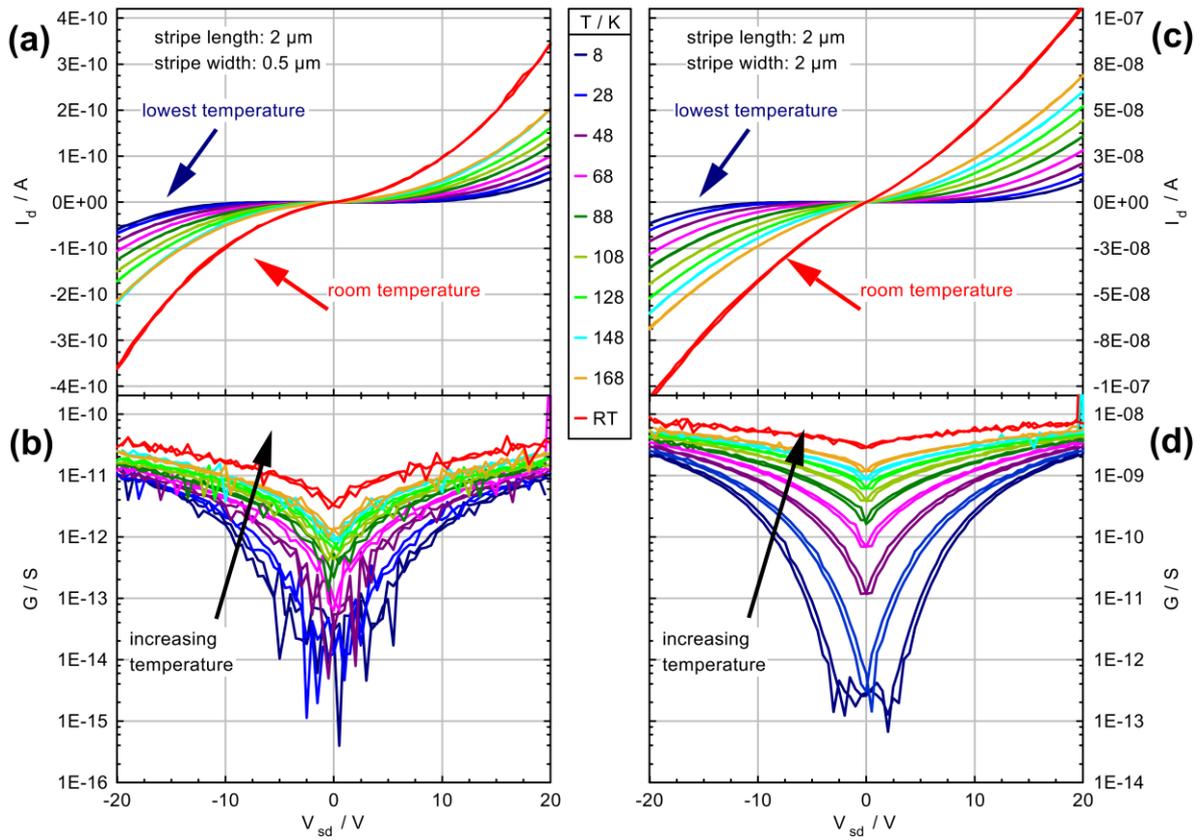

**Fig. 2** Output characteristics obtained from a CoPt nanoparticle stripe of 2 µm length and (a, b) 0.5 µm width as well as (c, d) 2 µm width at temperatures between 8 K and room temperature (RT). (a, c) Current data on a linear scale. (b, d) Differential conductances are displayed logarithmically for better visibility of the Coulomb-blockade regime, the corresponding threshold voltages, and the critical temperature. The back and forth sweeps show that there is virtually no hysteresis and demonstrate the reproducibility.

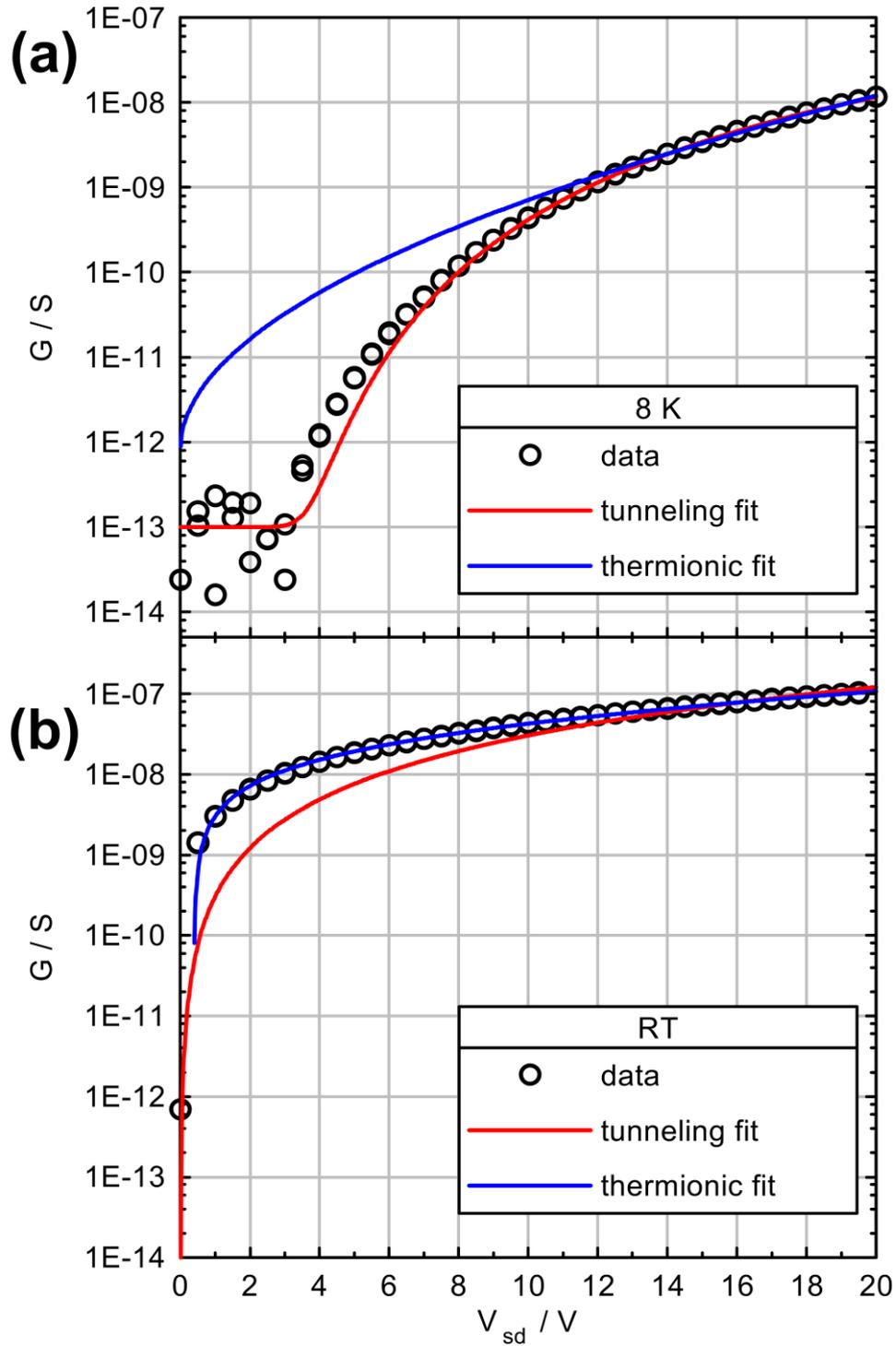

**Fig. 3** Different transport models are fitted to the obtained data from the CoPt nanoparticle stripe of 2.0 μm length and 2.0 μm width as reported in Fig. 2. (a) At 8 K, tunneling matches the current-voltage curves best, while (b) at room temperature the transport is mainly due to thermal contributions.

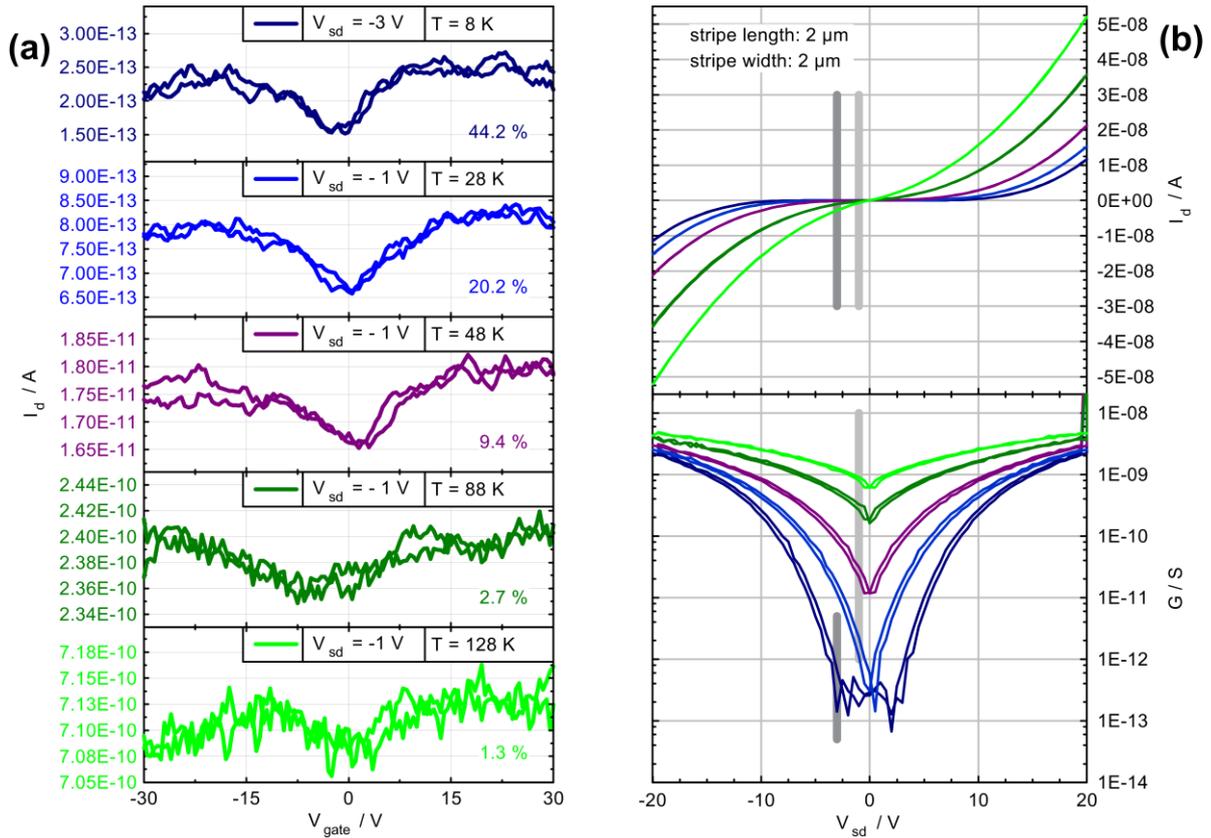

**Fig. 4** Temperature-dependent Coulomb oscillations obtained from the CoPt nanoparticle stripe of 2.0 µm length and 2.0 µm width as reported in Fig. 2. (a) Coulomb oscillations at the respective bias voltages for different measurement temperatures. The relative on/off-ratio $R_{\%}$, given next to the corresponding curve, decreases with increasing temperature. (b) S-shaped current-voltage curves and differential conductance plots for the exemplarily shown temperatures. The applied constant source-drain voltages are marked by vertical lines. The back and forth sweeps show that there is virtually no hysteresis and demonstrate the reproducibility.

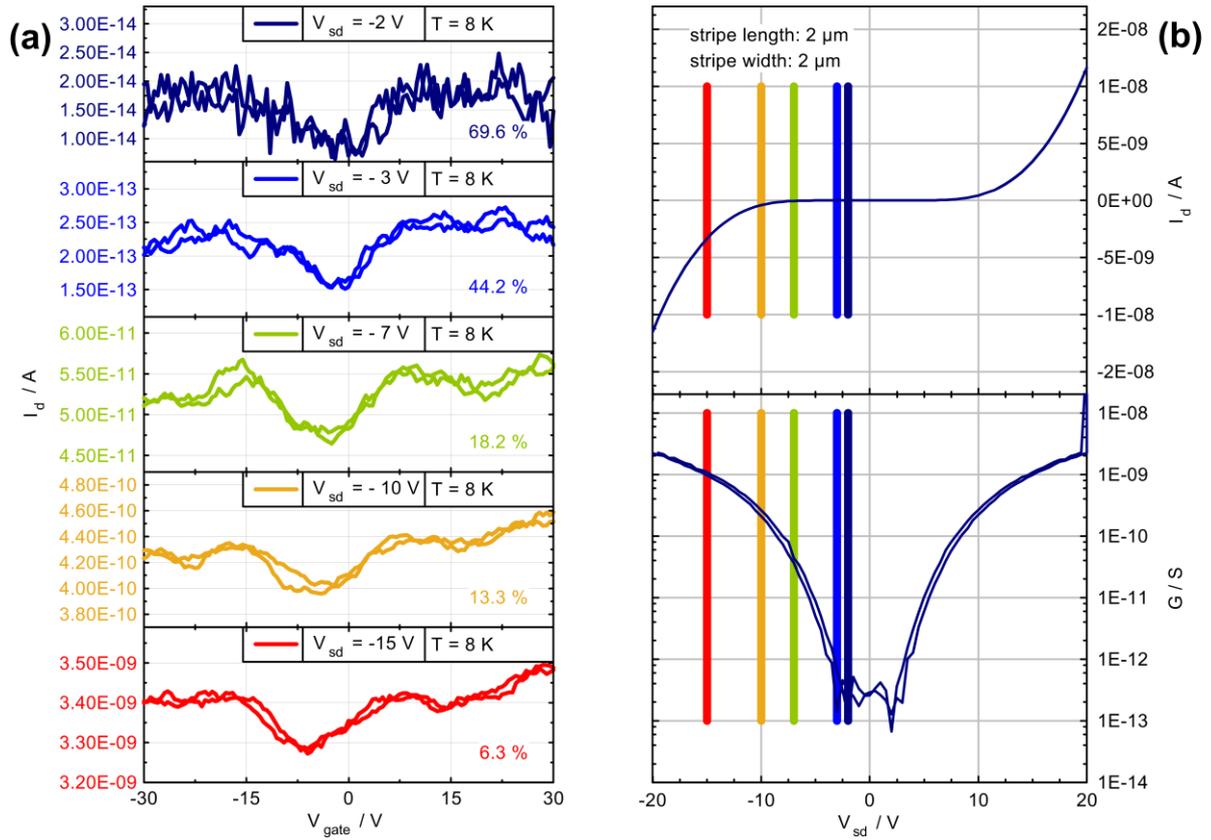

**Fig. 5** Coulomb oscillations at a temperature of 8 K for varying source-drain voltages obtained from the CoPt nanoparticle stripe of 2.0 µm length and 2.0 µm width as reported in Fig. 2. (a) Coulomb oscillations for different applied bias voltages. The relative on/off-ratio $R_\%$, given next to the corresponding curve, decreases with increasing bias voltage. (b) The applied source-drain voltages below, equal to, and above the threshold voltage are marked by vertical lines in the respective output characteristics. The back and forth sweeps show that there is virtually no hysteresis and demonstrate the reproducibility.